\documentclass[fleqn,usenatbib]{mnras}

\usepackage{newtxtext,newtxmath}

\usepackage[T1]{fontenc}

\DeclareRobustCommand{\VAN}[3]{#2}
\let\VANthebibliography\thebibliography
\def\thebibliography{\DeclareRobustCommand{\VAN}[3]{##3}\VANthebibliography}

\usepackage{graphicx}	
\usepackage{amsmath}	
\usepackage{mwe}
\usepackage{multicol}
\usepackage{ulem}
\DeclareUnicodeCharacter{2212}{-}
\usepackage[usenames,dvipsnames]{color} 

\title[Galaxy evolution in compact groups I]{Galaxy evolution in compact groups I: Revealing a transitional galaxy population through a multiwavelength approach}

\author[G.P. Montaguth et al.]{
Gissel P. Montaguth$^{1}$
\thanks{E-mail: gissel.pardo@userena.cl},
Sergio Torres-Flores$^{1}$,
Antonela Monachesi$^{1,2}$,
Facundo A. G\'omez $^{1,2}$, \newauthor
Ciria Lima-Dias$^{1,2}$,
Arianna Cortesi,$^{3}$,
Claudia Mendes de Oliveira $^{4}$,
Eduardo Telles$^{5}$, \newauthor
Swayamtrupta Panda $^{6}$,
Marco Grossi$^{3}$,
Paulo A. A. Lopes$^{3}$,
Jose A. Hernandez-Jimenez $^{7}$, \newauthor
Antonio Kanaan$^{8}$,
Tiago Ribeiro$^{9}$,
William Schoenell$^{10}$
\\
$^{1}$Departamento de Astronomía, Universidad de La Serena, Avda. Juan Cisternas 1200, La Serena, Chile\\
$^{2}$Instituto Multidisciplinario de Investigaci\'on y Postgrado, Universidad de La Serena, Ra\'ul Bitr\'an 1305, La Serena, Chile\\
$^{3}$Observat\'orio do Valongo, Universidade Federal do Rio de Janeiro, Ladeira Pedro Ant\^{o}nio 43, Rio de Janeiro, RJ, 20080-090, Brazil\\
$^{4}$Instituto de Astronomia, Geof\'isica e Ci\^encias Atmosf\'ericas da Universidade de S\~ao Paulo, Cidade Universit\'ria, CEP:05508-900, S\~ao Paulo, SP, Brazil \\
$^{5}$ Observat\'orio Nacional, Rua General Jos\'e Cristino, 77,  S\~ao Crist\'ov\~ao, 20921-400 Rio de Janeiro, RJ, Brazil\\
$^{6}$ Laborat\'orio Nacional de Astrof\'isica, R. dos Estados Unidos, 154 - Na\c{c}\~oes, Itajub\'a - MG, 37504-364, Brazil\\
$^{7}$ Universidade do Vale do Para\'iba, Av. Shishima Hifumi, 2911, S\~ao Jos\'e dos Campos, SP, 12244-000, Brazil\\
$^8$ Departamento de F\'isica, Universidade Federal de Santa Catarina, Florian\'opolis, SC, 88040-900, Brazil\\
$^9$NOAO, 950 North Cherry Ave. Tucson, AZ 85719, United States\\
$^{10}$ GMTO Corporation, N. Halstead Street 465, Suite 250, Pasadena, CA 91107, United States
}

\date{Accepted XXX. Received YYY; in original form ZZZ}
\pubyear{2023}
\begin{document}
\label{firstpage}
\pagerange{\pageref{firstpage}--\pageref{lastpage}}
\maketitle
\begin{abstract}
Compact groups of galaxies (CGs) show members with morphological disturbances, mainly products of galaxy-galaxy interactions, thus making them ideal systems to study galaxy evolution, in high-density environment. To understand how this environment affects the properties of galaxies, we select a sample of 340 CGs in the Stripe 82 region, for a total of 1083 galaxies, and a sample of 2281 field galaxies as a control sample. By performing a multi-wavelength morphological fitting process using S-PLUS data, we divide our sample into early-type (ETG), late-type (LTG), and transition galaxies using the $r$-band Sérsic index and the colour ($u-r$). We find a bimodal distribution in the plane of the effective radius-Sérsic index, where a  secondary ``peculiar'' galaxy population of smaller and more compact galaxies is found in CGs,  which is not observed in the control sample. This indicates that galaxies are undergoing a morphological transformation in CGs. In addition, we find significant statistical differences in the distribution of specific Star Formation Rate (sSFR) when we compare both environments for LTGs and ETGs. We also find a higher fraction of quenched galaxies and a lower median sSFR in CGs than in the control sample, suggesting the existence of environmental effects favoring the cessation of star formation, regardless of galaxy type. Our results support the notion that CGs promote morphological and physical transformations, highlighting their potential as ideal systems for galaxy pre-processing.

\end{abstract}

\begin{keywords}
galaxies: groups: general - galaxies: evolution - galaxies: interaction
\end{keywords}



\section{Introduction}

Interactions between galaxies play an important role in their evolution, generating changes in morphology (e.g., \citealt{1990Byrd, 2009Park},\citealt{2022Smith}), mass growth (e.g., \citealt{2011Cattaneo}), an increase of star formation (e.g \citealt{2013Patton} \citealt{Yoon_2020}) or cessation of star formation (e.g., \citealt{2008VandenBosch, 2013Woo}). Thus, the environment in which galaxies reside is a key ingredient in their evolution. Several studies have shown differences in the properties of galaxies that reside in different environments. For instance, elliptical and passive galaxies reside preferentially in regions of very high local density (\citealt{1984Dressler,2006Baldry}), and, the specific star formation rate is lower in denser environments compared to less dense ones (\citealt{2003Kauffmann}). \cite{2022Gonzales} found that the fraction of quenched galaxies is higher in groups of galaxies than in the field, and the fraction of quenched galaxies, both in groups and in the field, increases with stellar mass. They also found a larger fraction of blue quiescent and green valley galaxies in groups compared to the field. In addition, studies have shown that large numbers of galaxy cluster members have been pre-processed in groups and low-mass clusters (\citealt{2014A&A...570A.119E}, \citealt{2015Haines},\citealt{Bianconi2017}, \citealt{2019MNRAS.488..847P}, \citealt{2022Pallero}). Within this context, it is important to understand the physical processes that cause the pre-processing of galaxies in groups, in order to further understand the evolution of galaxies in different structures in the Universe.

Compact groups (CGs) of galaxies are ideal laboratories to study the effects of galaxy-galaxy interactions, mainly due to their high densities comparable to cluster cores, and  their low-velocity dispersions, $\sigma_G\sim200$km/s \citep{hickson1982systematic}. Indeed, these systems have from $3$ to $10$ luminous galaxies within a projected radius of the order of a few tens of kiloparsecs (\citealt{sohn2016catalogs}). One of the most famous sample corresponds to the Hickson Compact Group (HCGs) Catalog developed by \cite{hickson1982systematic}, and is the most studied sample of CGs, where he systematically selected 100 CGs from photometric plates of images. This environment contains a high fraction of galaxies that have morphological (\citealt{1994Claudia},  \citealt{2007Coziol}) or kinematic (\citealt{1991Rubin}, \citealt{2003Claudia}, \citealt{10.1093sergio}) peculiarities, which are typically associated to tidal interactions and mergers. Besides, when comparing CGs with less dense environments, such as the field or loose groups, it can be seen that CGs members have redder colours, which might indicate a higher fraction of early-type galaxies (\citealt{2004Lee}, \citealt{2008deng}, \citealt{2012Coenda}, \citealt{Poliakov_2021}). Galaxies located in CGs have preferentially larger concentration index, i.e. they are more compact, compared to systems located in less dense environments (\citealt{2008deng}, \citealt{2012Coenda}, \citealt{Poliakov_2021}). In addition, CGs have been studied to characterize the presence of intra-group light (e.g. \citealt{2005DaRocha}). For instance, \cite{Poliakov_2021} found that the surface brightnesses of the intra-group light component correlate with the mean CG morphology, such that brighter systems are dominated by early-type galaxies. Additionally, a gap or so-called ``canyon" has been observed in the mid-infrared (MIR) colour space for CG galaxies; galaxies with those colors are called infrared green valley galaxies, which are located between star-forming galaxies with red colours  and quiescent galaxies with blue colors in MIR. This feature is not detected in field galaxies (\citealt{2007AJohnson}, \citealt{2008Gallagher}), and it is only seen in a weaker form in galaxies falling in the Coma cluster (\citealt{2010Walker}). \cite{2013walker} found that galaxies in this MIR gap have already transitioned to the optical red sequence. 
This makes CGs an extreme environment for galaxy evolution, where the processes that transform galaxies from star-forming to quiescence act in an efficient way, and where photometric studies have been extremely useful.

The interaction history of galaxies in CGs can also be traced by studying their H\textsc{i} distribution. This allowed to propose an evolutionary scenario for CGs of galaxies which depends on the amount and distribution of neutral hydrogen. \cite{2001Verdes-Montenegro} found that HCGs members are deficient in H\textsc{i}, which may be produced by tidal stripping and/or heating. This lack of neutral gas affects the physical properties of galaxies in CGs. Furthermore, studies based on molecular gas \citep{2015bAlatalo} and neutral gas \citep{2023Jones} showed that only by knowing the amount of H\textsc{i} or H$_2$ gas, that galaxies in HCGs have, is not enough to infer their star formation activity. Indeed, H\textsc{i}-deficient galaxies still are active in infrared (i.e. star-forming galaxies can be detected in the infrared), suggesting that these galaxies, even though they lose their H\textsc{i}, still are able to maintain their star formation thanks to H$_2$ (\citealt{2023Jones}). On the other hand, \citet{2015bAlatalo} found that galaxies in HCGs do not need to eject their gas reservoirs in order to quench and undergo a transformation from blue spirals to red early-type galaxies. They also found that many of the galaxies in the "canyon" or infrared green valley galaxies contain molecular gas but cannot form stars efficiently, perhaps due to the existence of significant turbulence and shocks in the gas (\citealt{2015bAlatalo}), and most of these warm H$_2$ galaxies have suffered a significant decrease in molecular gas content and star formation (\citealt{2014Lisenfeld}) for HCG. Nevertheless, \cite{2016Bitsakis}, who used a larger sample, also found that shocks and turbulence have a relevant role in the cessation of star formation, and that the morphological transformation of late-types to earlier types occurs in the infrared green valley. All these studies indicate that physical and morphological processes in CG environments are quite complex. 

Within this context, it is still not clear what are the morphological differences between the CG galaxies and a sample of galaxies in a less dense environment, and how actually the CG environment favors morphological transformations. In addition, it is not clear whether there is any difference in the star formation of galaxies of the same morphological type in CGs compared to less dense environments. In other words,  does the CG environment promote cessation in the star formation of galaxies, i.e. are these systems responsible for the pre-processing of galaxies?

Given the relevance of CGs galaxies in the context of galaxy transformation and evolution, this study presents a comprehensive analysis of the physical and morphological characteristics of galaxies belonging to CGs. The primary data source for this analysis is the Southern Photometric Local Universe Survey (S-PLUS) project (\citealt{mendes2019southern}), which provides photometric information in 12 bands enabling a precise and detailed multi-wavelength analysis. We supplement this dataset with the GALEX-SDSS-WISE LEGACY catalogue (\citealt{2018GSWLC}). Through this work, we aim to systematically characterize and quantify the observed differences in galaxy properties between CGs and field galaxies. By achieving this in a standardized and homogeneous manner, our findings may help to enhance our understanding of the processes that drive galaxy evolution in CGs.

The outline of this paper is as follows: in Section \ref{sec:data}, we describe the data used and how we select the CGs as well as our control field sample. In Section \ref{sec:methodology}, we present the methodology for obtaining the structural parameters in a multi-wavelength analysis of each galaxy. In Sections \ref{sec:results} and \ref{sec:discussion} we present and discuss our results, and we summarize and conclude our work in section \ref{sec:conclusions}. Throughout this paper we have adopted a flat cosmology with $H_0 = 70 km$ $s^{-1}$ $Mpc^{-1}$, $\Omega_M = 0.3$, and $\Omega_\lambda = 0.7$ (\citealt{2003Spergel}).

\section{Data}
\label{sec:data}

In this Section, we describe the selection of the CGs of galaxies that we analyze, their photometric data as well as the control sample of field galaxies. We analyze a region inside the stripe data from the Stripe 82 region (\citealt{2009Dr7}) because it was associated with the first data release of the S-PLUS project (\citealt{mendes2019southern}). We note, however, that the analysis shown in this paper is based on the Data Release 3 (DR3) of S-PLUS, where the photometry has been improved. We complement these data with the GALEX-SDSS-WISE LEGACY catalog (\citealt{2018GSWLC}) to obtain Star Formation Rates (SFR) and stellar masses.

\subsection{Compact groups sample}
\label{sec:CGs}

To identify CGs in the S-PLUS catalog we use two catalogues to select the CGs, which cover a larger region of the sky because they are produced from SDSS data. The first one was published by \cite{sohn2016catalogs}, and it is built based on  a sample
of galaxies extracted from the SDSS-DR12. Groups were selected from photometric and spectroscopic information and using a Friends Of Friends code (FOF). Their CG selection is based on the following criteria: i) the absolute value of the average velocity of the group (V) minus the radial velocity of each member (V$_{i}$) should be lower than 1000 km s$^{-1}$; ii) each group must contain at least 3 members within $\Delta \text{mag}_r<3$ mag of the brightest member of the group; iii) galaxies should have a magnitude in $r$-band $<17.77$; iv) the CG should satisfy the compactness criterion, i.e. the surface brightness of the group, that is defined as the total magnitude of the group galaxies averaged over the smallest circle containing the galaxies (\citealt{hickson1982systematic}), should be lower than 26 mag arcsec$^{-2}$ in the $r$-band. The second catalogue was produced by \cite{zheng2020compact}, who used data from SDSS-DR14, LAMOST spectra, and GAMA survey. These authors select the CGs based on the combination of two methods: i) the \cite{hickson1982systematic} photometric criterion and ii) the spectroscopic technique, i.e. it uses the difference in radial velocities between the group and each individual galaxy. In this case each galaxy in a group must satisfy: a) richness and magnitude;  $3\leq N(14.00\leq r\leq17.77)\leq10$, b) isolation; $\theta_N\geq3\theta_G$, where $\theta_G$ is the radius of the group and $\theta_N$ is the radius of a bright galaxy closest to the group, c) compactness; $\mu_r\leq26.0$ mag arcsec$^{−2}$, and d) velocity difference; $|V-V_{i}|\leq1000$ km/s. The second catalogue contains a greater number of CGs, although there are some CGs that appear in both catalogues, with $46\%$ of the CGs that are in the \cite{sohn2016catalogs} catalogue also listed in the \cite{zheng2020compact} catalogue. However, the \cite{sohn2016catalogs} complements the \cite{zheng2020compact} catalogue, especially at low redshift, since the latter catalogue contains galaxies brighter than 14.0 mag (petrosian magnitude in the $r$-band), whereas there is no bright limit applied on the magnitudes of the galaxies in the \cite{sohn2016catalogs} catalogue. It should be noted that these two catalogues do not have any colour selection.
The combination of these two catalogues produces an initial galaxy position and redshift catalogue of 424 CGs, which will be cross-match with S-PLUS, as described below.

\subsubsection{S-PLUS Photometric data}

S-PLUS is an ongoing imaging survey that began its observations in 2016 and will cover a region of  $\sim 9300$ $deg^2$ on the sky, using a robotic $0.8$m aperture telescope at the Cerro Tololo Inter-American Observatory (CTIO), Chile. S-PLUS uses the Javalambre 12-band magnitude system (\citealt{2019Cenarro}), which includes the 5 broad-band filters $u$, $g$, $r$, $i$, $z$ and 7 narrow-band filters centered on prominent stellar spectral features: the Balmer jump/[OII], Ca H$+$K, H$\delta$, G-band, Mg b triplet, H$\alpha$, and the Ca triplet ($J0378$, $J0395$, $J0410$, $J0430$, $J0515$, $J0660$ and $J0861$) (\citealt{mendes2019southern}). S-PLUS is divided into five sub-surveys and the Stripe 82 region belongs to the main survey. Stripe 82 comprises the coordinates between $0^{\circ}$ to $60^{\circ}$ and $300^{\circ}$ to $360^{\circ}$ in right ascension and $-1.4^{\circ}$ to $1.4^{\circ}$ in declination. More details about the survey can be found in \cite{mendes2019southern}. The exposure times of the main survey are 681, 660, 354, 177, 171, 99, 183, 120, 870, 138, 240, and 168 seconds for filters $u$, $J0378$, $J0395$, $J0410$, $J0430$, $g$, $J0515$, $r$, $J0660$, $i$, $J0861$ and, $z$, respectively. 

The S-PLUS data were reduced by using the \textit{jype} pipeline designed for J-PLUS and J-PAS (\citealt{2015hsa8.conf..798C}). \cite{almeida2021data} explained in detail the procedure to generate the S-PLUS catalogues. The source detection and photometry are done with the SExtractor software (\citealt{bertin1996sextractor}) and the zero-point photometric calibrations were estimated by using an optimized technique for wide-field multi-filter photometry\footnote{The zero points are available in the auxiliary tables section \url{https://splus.cloud}}. The photometric depths of S-PLUS were defined for different values of S/N, where the lowest value is S/N=3. Then, the limiting magnitude is defined as the peak of the magnitude distribution at that S/N. In the case of a S/N=3, the shallowest filter is $J0861$ with a petrosian magnitude of $19.9$. The deepest magnitudes are reached in filters $g$ and $r$ with $21.3$ (\citealt{almeida2021data}). Reaching surface brightness of $\sim 24.5$ mag arcsec$^{-2}$ in the $r$-band.

The S-PLUS DR3 catalogue contains astrometric, structural, and photometric information for each source, such as the celestial coordinates (RA, DEC) in the J2000 system, the physical position on the CCD (X, Y), the size of the different photometric apertures, their magnitudes and errors, the signal to noise ratio, the FWHM, the parameters on the isophotes (A, B, and THETA), the light fraction radii (FLUXRADIUS). The different apertures included were "AUTO", "PETRO", "ISO", "APER3", "APER6", "PSTOTAL", where "APER3" and "APER6" are circular apertures of 3 and 6 arcsec diameters, respectively. The magnitudes in the S-PLUS catalogue have not been corrected for Galactic extinction. In this work, we use the information on the RA and DEC positions, the AUTO and APERS3 magnitudes in the twelve filters, the physical position in the CCD, and the parameters on the isophotes of the sources in CG galaxies from the S-PLUS catalogs in DR3.

Our sample of 424 CGs (described in Section \ref{sec:CGs}) was matched with the S-PLUS DR3 catalogues. We find 340, out of the 424, CGs in the S-PLUS database ($80\%$ of the CG sample). These CGs had all their galaxies identified in S-PLUS, with 1092 galaxies at redshifts $0.015<z<0.197$.  We eliminated 9 galaxies from the total sample because they were visually contaminated by bright stars. In total, we have 1083 galaxies. This is our final sample for CGs galaxies that will be analyzed throughout this paper. 

\subsubsection{Achival GALEX-SDSS-WISE LEGACY photometric data}

In order to complement our analysis, we search for multi-wavelength data in the \textit{GALEX}-\textit{SDSS}-\textit{WISE} LEGACY (GSWL) catalogue, published by \cite{2018GSWLC}. We find that $88\%$ of our selected CG galaxies are included in this catalogue, corresponding to 321 CGs with a total of $967$ galaxies. \cite{2018GSWLC} fit the Spectral Energy Distribution (SED) for galaxies in the GSWL catalogue by using the CIGALE code (\citealt{2009Noll}) and considering the \textit{GALEX}, \textit{SDSS} and \textit{WISE} photometry data calibrated in Herschel-ATLAS. In this work, we will use the SFR derived from their SED fittings to complement our analysis in the following sections.\\ 

\begin{figure}
    \centering

    \includegraphics[width=\columnwidth]{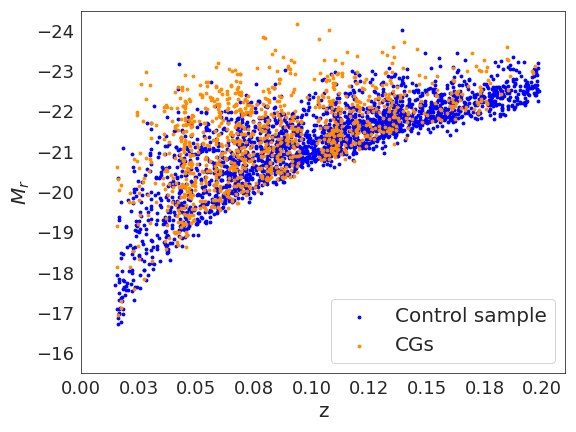}
    \includegraphics[width=\columnwidth]{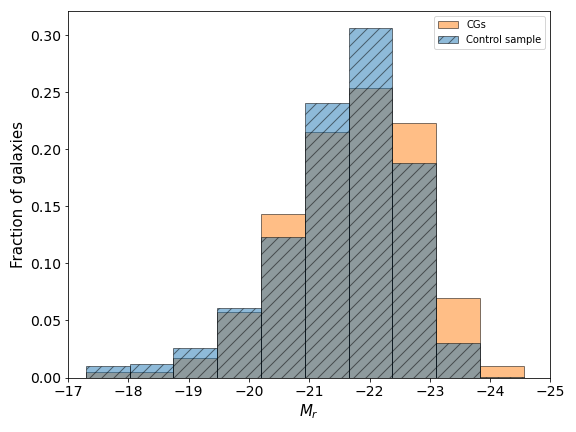}

    \caption{The upper panel shows the absolute magnitude in $r$-band as a function of redshift, and the lower panel shows the normalized histogram distribution of M$_r$ in both environments. Galaxies in CGs are in orange whereas galaxies in the control sample are shown as blue points. Absolute magnitudes in $r$-band corrected for galactic extinction and with K-correction.}
    \label{fig:completes}
\end{figure}

\subsection{Control sample: Field galaxies} 

In order to determine the influence of the environment on the properties of CG galaxies (e.g. size, colours, and SFRs), we have defined a sample of galaxies located in less dense environments, namely field galaxies, which we use as our control sample. We selected this sample to have the same apparent magnitude in $r$-band and redshift limits as the sample of galaxies in CGs.

The field galaxies are selected from the catalogue published by \cite{2007Yang} as those labeled as \textit{single-member groups} (i.e. a halo of an individual galaxy) in the Stripe 82 area. \cite{2007Yang} used spectroscopic data from SDSS-DR4 (\citealt{2006SDSSDR4}) and NYU-VAGC (\citealt{2005NYU-VAGC}) and a similar method to the group finder used by \cite{Yang05}. 

We cross-match the field sample with the S-PLUS DR3 catalogue obtaining a sample of $11841$ galaxies for the same redshift interval at which we have our CGs sample, $0.015<z<0.197$. From this sample, we randomly select a subsample of $2300$ galaxies using a Monte Carlo algorithm which preserves the characteristics of the initial sample. We consider from now on only this sample of $2300$ control field galaxies for our analysis. We excluded 19 galaxies from this sample because they either suffered from contamination caused by saturated stars affecting their flux or exhibited low surface brightness, i.e., they are diffuse. For this control sample of $2281$ galaxies, we also perform a match with the GSWL catalogue, finding $92\%$ of galaxies in this catalogue.

\subsection{Final sample of galaxies}

The top panel of Figure \ref{fig:completes} shows the $r$-band absolute magnitude of galaxies as a function of redshift. We estimate the absolute magnitude from $M_r=m_r-5\times log(D_L/10pc)-K$. Here, $D_L$ is the luminosity distance calculated from the redshift and $K$ is the K-correction (\citealt{Chilingarian2010}) \footnote{The K-correction was made with the code that is available at\url{http://kcor.sai.msu.ru/}}, used to transform the observed magnitudes into magnitudes in the rest frame. All the galaxies in our sample have observed (g-r) colors within the range of -0.1 to 1.9,  which ensures that we can apply the K-correction by \citet{Chilingarian2010}. Blue points in the figure are galaxies in our control sample and orange points are galaxies in the CGs, as described in \ref{sec:CGs}. 
 Magnitudes have been corrected for Galactic extinction by using \cite{1989Cardelli}, extinction law with $R_v=3.1$, and the maps from \cite{2011schlafly}.   
Although in S-PLUS it is possible to reach magnitudes as faint as $r$-band$\sim21$, our sample is limited to galaxies spectroscopically selected from the Main Galaxy Sample catalogue in the legacy SDSS. This catalogue is complete for a magnitude of $r \leq 17.77$ (petrosian magnitude corrected for Galactic Extinction). We note that the sample chosen by \cite{2007Yang},  \cite{sohn2016catalogs}, and \cite{zheng2020compact} used the SDSS r-band magnitude for selecting their samples, whereas in our case, we employed S-PLUS data for the photometric measurements. This difference may account for the minor dispersion observed at the bottom region in the $z$ vs $M_r$ plane (top panel in Figure \ref{fig:completes}).

In the bottom panel of Figure \ref{fig:completes} we show that the sample covers absolute magnitudes which range between $M_{r}=-17$ to $M_{r}=-24$, with a peak at $M_{r}=-22$. Since our goal is to study how the environment impacts galaxy evolution, we study all galaxies detected in the CGs, and compare them with a similar magnitude distribution in the field sample, as shown in Figure \ref{fig:completes}.

\section{Methodology}
\label{sec:methodology}
\subsection{Morphometric parameters} \label{subsec:galfitm}

The use of photometric information in 12 bands allows us to study the morphology of galaxies in different wavelength ranges. To obtain the morphometric parameters of each galaxy we use the MegaMorph  code (\citealt{2011Bamford}, \citealt{2013Haussler},\citealt{2013vika}) which performs a two-dimensional fitting at multiple wavelengths by using the GALFITM algorithm, a modified version of GALFIT 3.02 (\citealt{2002Peng}, \citealt{2010Peng}). This code models the surface brightness of a galaxy by using a two-dimensional analytical function, that can be a Sérsic (\citealt{Sersic}), Nuker (\citealt{1995lauer}), de Vaucouleur (\citealt{1948Vaucouleurs}) or Exponential (\citealt{1970Freeman}). Additionally, to extend these functions to multiple wavelengths, free parameters are replaced by wavelength functions, which are a set of Chebyshev polynomials. GALFITM performs multiple fitting components where users can include information regarding the sky background, as well as decompose the galaxy in bulge, disk, and bar. The advantage of using a multi-wavelength morphological fitting over single-band fitting is that it improves the results in terms of accuracy and robustness (\citealt{2013vika}). The best fit is determined using the Levenberg-Marquardt (LM) algorithm, minimizing $\chi^2$.

All galaxies were fitted with a single component model, following a Sérsic profile, as shown in equation \ref{eq:sersic}:

\begin{equation}
    \centering
    I=I_e \exp{[-b_n ((\frac{R}{R_e})^{1/n}-1)]}
\label{eq:sersic}
\end{equation}

where $I_e$ is the intensity at the effective radius $R_e$ (i.e. the radius containing half of the total light), $b_n$ depends on $n$ as $\Gamma(2n)= 2\gamma(2n, bn)$, where $\Gamma$ and $\gamma$ are the Gamma function and lower incomplete Gamma function (\citealt{1991Ciotti}), respectively, and $n$ is the Sérsic index which determines the shape of the light profile. For instance, $n = 1$ typically represents an exponential profile for galactic disks (\citealt{1970Freeman}), and $n = 4$ is associated with a de Vaucouleurs profile (\citealt{1948Vaucouleurs}), which is usually associated with massive spheroidal components such as elliptical galaxies or galaxy bulges.

Since the surface brightness profiles are simultaneously fitted in all filters, MegaMorph requires all the S-PLUS multi-wavelength images as input parameters. It also needs a file with initial parameters such as photometric zero points and seeing, and the point spread function (PSF) derived for each band. In all cases, the PSF was modeled with a Moffat function, $PSF(r)=\frac{\beta-1}{\pi \alpha^2} \left[1+ \left(\frac{r}{\alpha}\right)^2\right]$ and $FWHM=2\alpha \sqrt{2^{1/\beta}-1}$ where the full-width half maximum (FWHM) and beta parameter ($\beta$) are available in the DR3 catalogues.

\begin{figure*}
    \includegraphics[scale=0.44, angle=90]{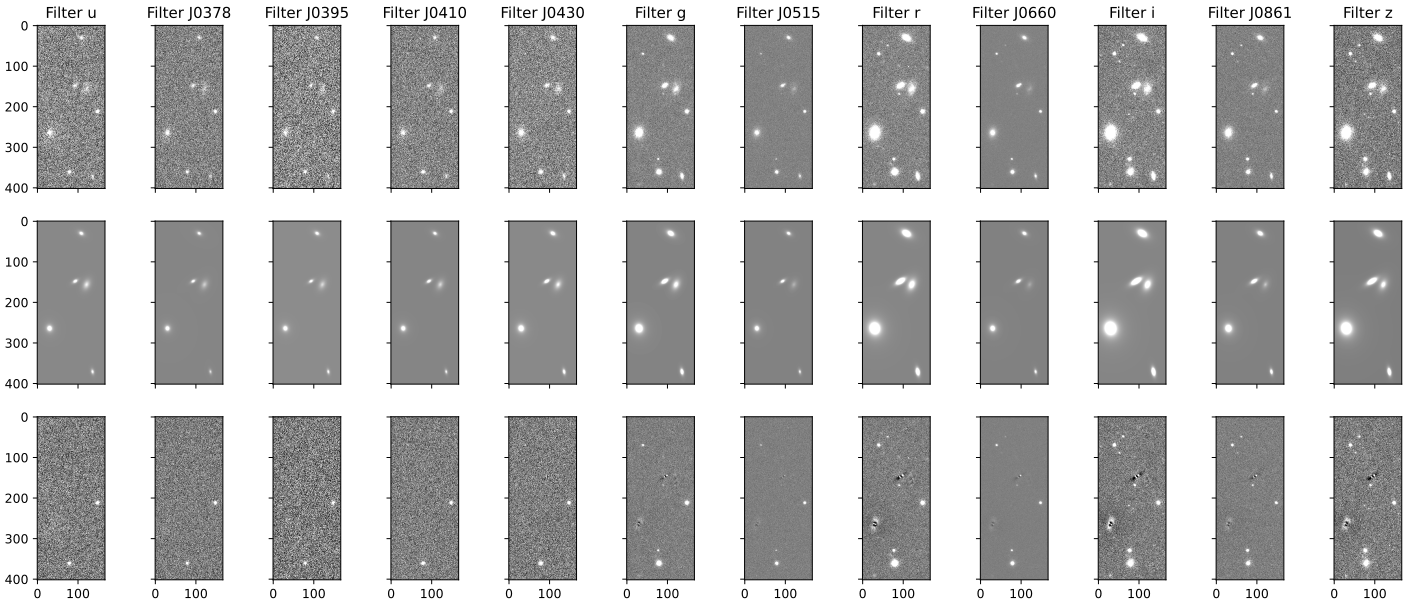}
    \caption{Left panels: S-PLUS images for one of the CGs analyzed in each filter. Middle panels: the model generated using GALFITM. Right panels: the residual, the model subtracted from the observed image. The box size of each plot corresponds to $220"\times 92"$.}
   \label{fig:215}
\end{figure*}

In addition, we masked some images in order to avoid spurious detections produced by foreground/background sources in the line of sight of CGs galaxies. To obtain the masks we first generate the segmentation images using the SExtractor software \citep{bertin1996sextractor}, which masks all the sources it detects, including the galaxies in the CGs we are interested in.  As a second step, we use the generated images to create a new segmentation image, where we unmask the galaxies of the CGs. We then assign a numerical value of one to the pixels that are masked, and a value of zero to those that are not; this is because GALFITM requires that the mask we provide as input must contain zeros for the regions to be fitted. Figure \ref{fig:215} shows an example of a GALFITM output for one of the CGs analyzed in this work. In this figure, the top panels display the images in each S-PLUS filter, the middle panels show the light profile model, and the bottom panels show the residual image, i.e. the model subtracted from the observed image.

\subsection{Stellar masses estimation}

\cite{2001Bell} and \cite{2011Taylor} have demonstrated that a robust and reliable estimate of the stellar mass of a galaxy can be obtained from optical bands, thanks to the mass-luminosity relation, which is derived from stellar population synthesis models. \cite{2011Taylor} used the Single Stellar Population (SSP) models by \cite{2003Bruzual} and assumed a \cite{2003Chabrier} Initial Mass Function. In order to correct for the internal extinction of galaxies they use the extinction law proposed by \cite{2001PASP..113.1449C}. They derived the following relation to estimate $M_*/L$ using the colour $(g-i)_0$, at rest frame, and the absolute magnitude in the $i$-band, $M_i$:

\begin{equation}
    log_{10}(M_*/M_{\odot}) = 1.15+0.7\times(g-i)_0-0.4\times M_i
\label{eq:masa}
\end{equation}

\begin{figure}
    \includegraphics[width=\columnwidth]{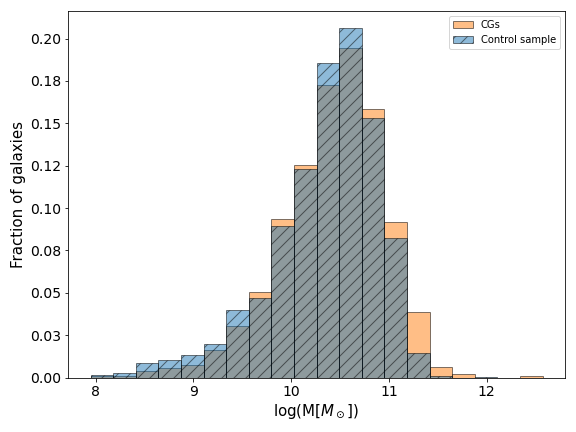}
    \caption{Normalized histograms of stellar masses for galaxies in CGs (orange) and in the control sample (blue).}
    \label{fig:masses}
\end{figure}

We use Equation \ref{eq:masa} to estimate the stellar mass  using the S-PLUS data of the galaxies in our samples, both in the control sample and in the CGs. Figure \ref{fig:masses} shows the histograms of the stellar masses in both environments, in blue and orange for the galaxies in the control sample and in CGs, respectively.

\begin{figure}
    \centering
    \includegraphics[width=\columnwidth]{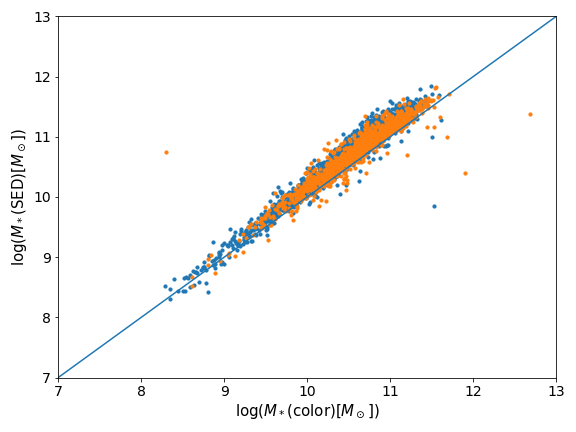}
    \caption{Comparison between the stellar masses estimated in this work and that estimated in the GSWL catalog using SED fitting. Orange points represent galaxies in CGs and the blue ones galaxies in the control sample. The  solid line is the one-to-one relationship between the compared parameters.}
    \label{fig:comparacion}
\end{figure}

In Figure \ref{fig:comparacion} we compare the stellar masses that we estimate with those published in the GSWL catalogue (based on SED fitting), where the blue and orange dots represent the galaxies in the control sample and the CGs, respectively. The solid line represents a one-to-one relationship. The data follow closely this relation with some scatter, which indicates that using the colour $(g-i)$ and the Equation \ref{eq:masa} proposed by \cite{2011Taylor}, provides a good estimate of stellar mass. We thus use the stellar masses obtained from Equation \ref{eq:masa} throughout this paper.

\subsection{Star Formation Rates}

The star formation rates (SFR) that we use in this work are those estimated by \cite{2018GSWLC}, based on the SED fitting done using the CIGALE code as mentioned in Section 2.1.2. Those authors assumed a two-component exponential star formation history, one for the younger population and another for the older one. The formation times for the younger component range from 100 Myr to 5 Gyr, where the mass fraction of the younger component must range up to 0.5. The old population times ranged from 850 Myr to 20 Gyr. Using the SFR from GWLS catalogue and the stellar mass estimated by us in the previous section, we estimate the specific star formation rate as:

\begin{equation*}
    sSFR=\frac{SFR}{M_*} [yr^{-1}]
\end{equation*}

\section{Results}
\label{sec:results}

We explore in this section the relation between the structural and physical parameters of the galaxies analyzed in this work.

\subsection{Structural parameters}
\label{sec:Structural}

\subsubsection{Classifying galaxies: Late-type, transition, and early-type systems} 
\label{sec:class}

The colours of galaxies are associated with their predominant stellar populations and are related to their morphology (\citealt{morgan1957spectral}). In particular, blue galaxies are typically star-forming objects, whereas red galaxies are mostly quiescent, containing red and old stars. Indeed, we may expect a correlation between colours and structural parameters of galaxies (e.g. \citealt{vika2015megamorph}). As indicated in section \ref{subsec:galfitm}, in this work we characterize the morphology of each galaxy by fitting its light profile with the Sérsic function. Therefore, colours and Sérsic index can be used to separate early-type galaxies (ETGs) from late-type galaxies (LTGs), where red galaxies with high values of $n$ are classified as ETGs and blue galaxies with low values of $n$ are classified as LTGs (\citealt{2008Ball}, \citealt{Kelvin2012}, \citealt{vika2015megamorph}).

\begin{figure}
    \includegraphics[width=\columnwidth]{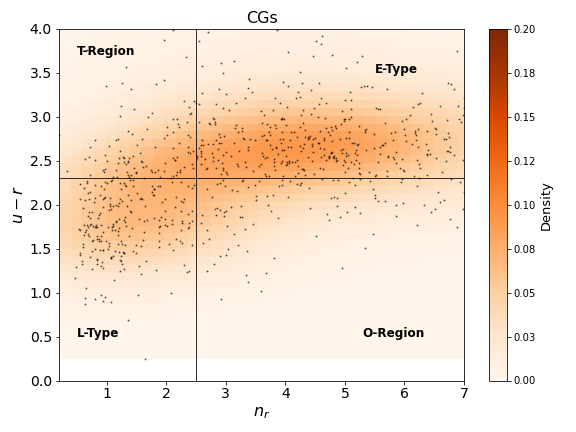}
    \includegraphics[width=\columnwidth]{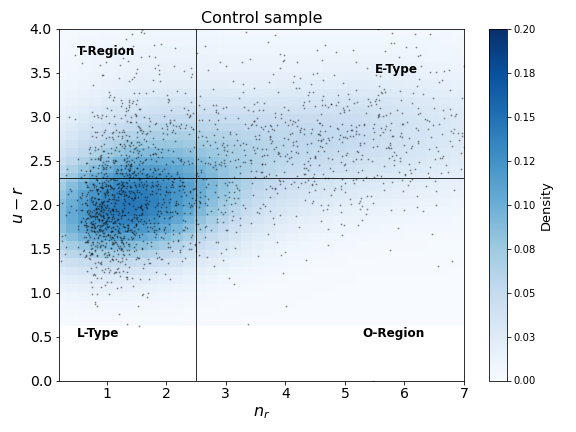}
  \caption{ETGs, transition galaxies, and LTGs classification, using $(u-r)$ colour and Sérsic index in the $r$-band $(n_r)$, the vertical line is for $n_r= 2.5$ and horizontal for $(u-r) = 2.3$. The upper plot corresponds to the galaxies in CGs and the lower plot to the galaxies in the control sample, in blue and orange are density plots, where the black dots are the original data.}
  \label{fig:E_T_population}
\end{figure}

We use the Sérsic index in the $r$-band and the $(u-r)$ colour to classify galaxies as ETG and LTG following the criteria proposed by \cite{vika2015megamorph}.  Galaxies with $n\geq2.5$ and $(u-r)\geq2. 3$ are ETGs and galaxies with $n<2.5$ and $(u-r)<2.3$ are LTGs. The top panel of Figure \ref{fig:E_T_population} shows this classification for galaxies in the CGs. The same classification for our control sample of field galaxies is shown in the bottom panel of Figure \ref{fig:E_T_population}. The galaxies located in the top-left region of each  $n-$colour diagram, i.e. red colours and $n<2.5$, are defined as \textit{transition galaxies}. Previous authors have named this region as red low-$n$ galaxies (\citealt{2014vulcani}) or red disk galaxies, where they used colour and concentration to classify them or just the colour and the morphological classification from Galaxy Zoo (\citealt{lopes2013},\citealt{Tojeiro2013}). In the Appendix \ref{sec:apen_Other} we present the results we find for the region with $n>2.5$ and $(u-r)<2.3$, which we have defined as \textit{Other region}. For \cite{2014vulcani} the galaxies in this region are a mixture of green and blue high-$n$.

In Table \ref{tab:resumen} we list a summary of the median values for the different physical and morphological parameters obtained for the  populations of ETGs, LTGs, and transition galaxies, classified as described above (see Fig. \ref{fig:E_T_population}). The values outside and inside the parenthesis represent the CGs and the control sample, respectively. We find that the median values of the effective radius and stellar mass are lower in CG galaxies, except for LTGs. The Sérsic index is similar for LTGs in both environments, however, it is slightly higher for transition galaxies in CGs, whereas for ETGs this value is slightly higher in the control sample. In addition, The median of $sSFR$ is lower in CGs than in the control sample regardless of the morphological type. Furthermore, in CGs, the percentage of quenched galaxies is higher than in the control sample for all three types of galaxies, where we consider a galaxy to be quenched if $Log(sSFR)\leq-11$, based on the criterion proposed by \cite{2013Wetzel}. The implications of these results are discussed in the following sections

\begin{figure}
    \centering
    \includegraphics[width=\columnwidth]{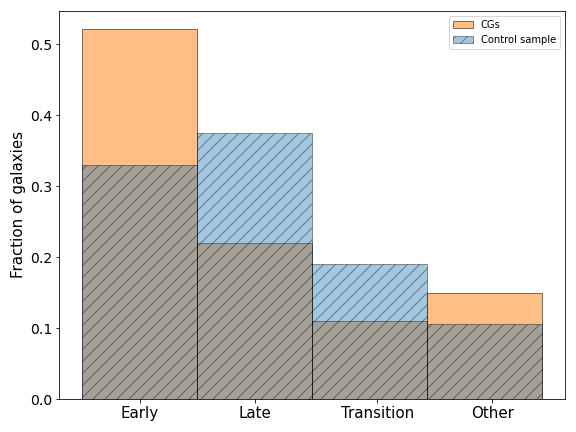}
    \caption{Histogram of the fraction of galaxies in each population, LTGs, ETGs, transition galaxies, and galaxies in the right lower region (other), defined in the text and shown in figure \ref{fig:E_T_population}.}
    \label{fig:hist_el}
\end{figure}

\begin{figure}
    \centering
    \includegraphics[width=\columnwidth]{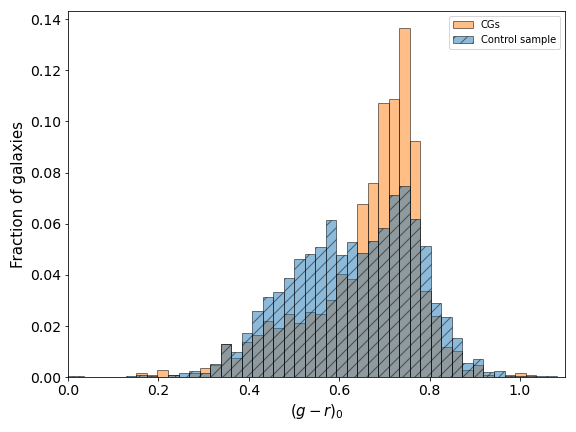}
    \caption{Histograms of the colour distribution ($g-r$), for both environments.}
    \label{fig:color}
\end{figure}

Following this classification, and as can be seen in the histogram shown in Figure \ref{fig:hist_el} and in the first row in Table \ref{tab:resumen}, there is a larger fraction of ETGs in the CGs than in the control sample, while in the control sample, there is a larger fraction of LTGs and transition galaxies. This  suggests an environmental difference in the fraction of ETGs and LTGs between the CGs and the control sample. However, it is worth noting that this may be due to a selection effect in the CGs sample since one of the criteria used is that they should have surface brightnesses $\mu_r\leq26.0$ $mag$ $arcsec^{−2}$, which favours groups with bright galaxies such ETGs. In addition, the catalogue of CGs used in this work was produced by using optical data, which likely generated a bias toward groups with massive galaxies (\citealt{2015Hernandez}). This is also reflected in the colour distribution of these galaxies. 
Figure \ref{fig:color} shows the histogram of the $(g-r)_0$ colours at rest frame. CG galaxies display a bimodal colour distribution with a red component that dominates, which peaks at a colour $(g-r)_0\sim 0.75$, with a small tail at bluer colours, while galaxies belonging to the control sample show a clear bimodality.

\begin{table*}
\begin{tabular}{|l|l|l|l|}
\hline
                                              & \textbf{Early type}                                      & \textbf{Transition region}                               & \textbf{Late type}                                        \\ \hline
\textbf{Percentage}                           & $52\%$ ($33\%$)                                    & $11\%$ ($19\%$)                                    & $22\%$ ($37\%$)                                     \\ \hline
\textbf{Median $R_e$[Kpc]${[}r{]}$}              & $7.24^{7.70}_{6.44}$ ($8.77^{9.34}_{8.04}$)         & $5.78^{6.69}_{4.78}$ ($7.01^{7.37}_{6.71}$)      & $5.31^{5.73}_{4.94}$  ($6.56 ^{6.84}_{6.31}$)        \\ \hline
\textbf{Median $n{[}r{]}$}                   & $4.64 ^{4.78}_{4.52}$  ($4.94 ^{5.11}_{4.78}$)      & $1.61^{1.83}_{1.46}$  ($1.41 ^{1.46}_{1.38}$)       & $1.19^{1.25}_{1.10}$  ($1.13^{1.18}_{1.10}$)         \\ \hline  
\textbf{Median $Log(M_*[M_{\odot}])$}         & $10.70^{10.75}_{10.67}$  ($10.77 ^{10.79}_{10.74}$) & $10.51^{10.61}_{10.44}$  ($10.63 ^{10.68}_{10.60}$) & $10.07 ^{10.12}_{10.02}$  ($10.18 ^{10.21}_{10.14}$  ) \\ \hline
\textbf{Median $Log(sSFR[$yr$^{-1}])$} & $-11.99^{-11.91}_{-12.07}$  ($-11.92^{-11.86}_{-12.00}$)  & $-11.00^{-10.75}_{-11.07}$  ($-10.83 ^{-10.73}_{-10.97}$) & $ -10.16^{10.11}_{-10.23}$  ($-10.07^{-10.10}_{-10.03}$  )  \\ \hline
\textbf{Percentage of quenched galaxies}                    & $87\%$ ($82\%$)                              & $50\%$ ($45\%$)                              & $ 10\%$    ($3\%$)                            \\ \hline 
\end{tabular}
\caption{Median values of the different physical and morphological properties derived for each galaxy population for CGs. The values in parentheses correspond to those of the control sample. The supra and subscript are the values at the 90\% confidence interval estimated with the bootstrapping technique for each median. We considered quenched galaxies as those with $Log(sSFR)\leq-11$.}
\label{tab:resumen}
\end{table*}

\subsubsection{Sérsic index behaviour as a function of wavelength}
\label{sec:n_wave}

\begin{figure*}
    \centering
    \includegraphics[width=\columnwidth]{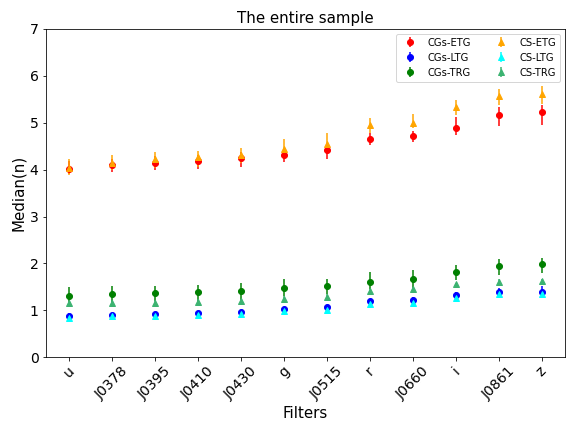}
    \includegraphics[width=\columnwidth]{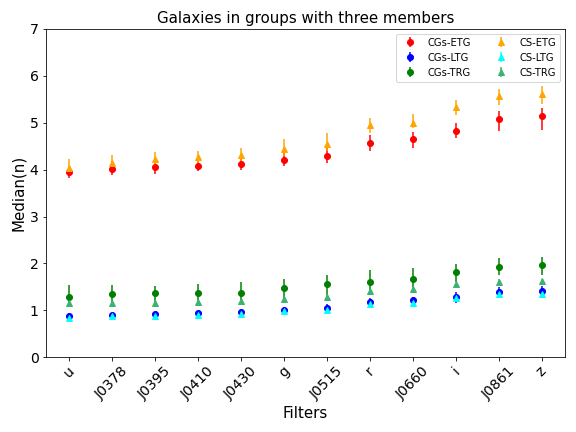}
    \includegraphics[width=\columnwidth]{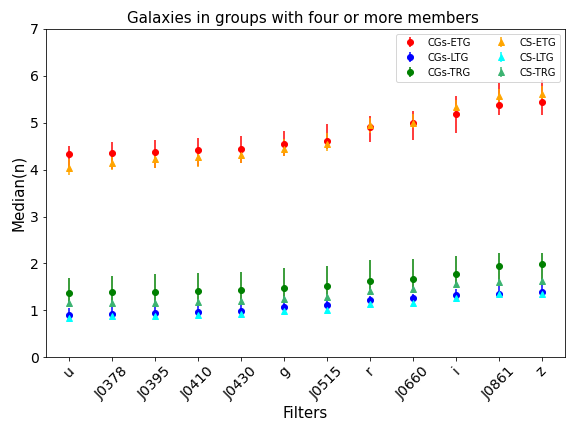}
    \caption{The median Sérsic index ($n$) as a function of S-PLUS filters for ETGs, LTGs, and transition galaxies for the CGs (circles) and the control sample (triangles). The error bars are the 90\% confidence interval using bootstrapping. The upper left plots contains the entire sample in both environments, the top right plot for CGs contains the galaxies in groups with three members, and the lower panel only the galaxies in CGs in groups with 4 or more members.}
    \label{fig:n_fil}
\end{figure*}

Since the Sérsic index is available for each of the 12 S-PLUS filters, we can analyze the median behaviour of this parameter with respect to wavelength, for each population in each environment. 
In Figure \ref{fig:n_fil} we show the median Sérsic index that we obtain for each galaxy population, as a function of wavelength. The triangles and circles in Figure~\ref{fig:n_fil} represent the median Sérsic index for the galaxies in the control sample and in CGs, respectively. LTGs, transition galaxies and ETGs (as described in section \ref{sec:class}) are shown as blue/cyan, green/light green, and red/orange colours, respectively. Uncertainties were estimated from a bootstrapping technique with a confidence interval (CI) of 90\%. Top left, top right, and bottom panel show the entire CGs sample, CGs that contain three members, and CGs that contains four or more members, respectively. This separation is made because \cite{hickson1982systematic}, originally proposed that the CGs should have at least 4 members. Given that the Sérsic index is one of the most important parameter in this work, we want to check if there are differences in our results if our sample is subdivided according to the members the CGs have. We note, however, that \cite{2013Duplancic} showed that galaxy triplets satisfying the other CGs selection criteria did not differ in terms of the total star formation activity and global colours from more populated CGs. We find, as it can be seen in Fig.~\ref{fig:n_fil}, that there is no difference in the results presented here (Sérsic index as a function of wavelength) when subdividing the sample of CGs. For this reason, we decide that for the following analyses, we would use all galaxies in CGs without discriminating the number of members.

In general, we find that the median of $n$ increases with wavelength. In the case of ETGs in CGs (in the control sample) the value of $n$ increases from the bluest filter to the reddest by $23\%_{16\%}^{29\%}$ ($28\%_{22\%}^{34\%}$). For galaxies in the transition region $n$ increases by $34\%_{17\%}^{49\%}$ ($29\%_{21\%}^{35\%}$). For LTGs we find that $n$ increases by $37\%_{25\%}^{48\%}$ ($38\%_{33\%}^{44\%}$). Also, we note that ETG in CGs exhibit slightly lower $n$ at redder filters, than those of the control sample. On the contrary, in the transition region, the increase of the median value of $n$ is higher for the CG galaxies, especially in the red filters. This is between $\sim 10\%$ and $\sim 28\%$ for the median $n$ in CGs compared to the control sample. Section \ref{sec:SFR} will analyze the transition galaxies in more detail, connecting it with other properties such as SFR, $R_e$, and $n$.

\subsubsection{Effective radius as a function of mass, Sérsic index, and wavelength}
\label{sec:re}

\begin{figure*}
    \centering
    \includegraphics[scale=0.34]{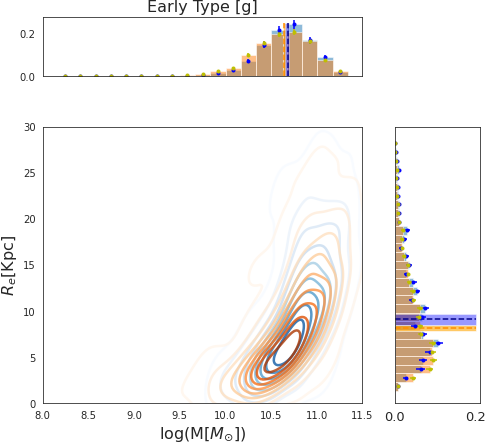}
    \includegraphics[scale=0.34]{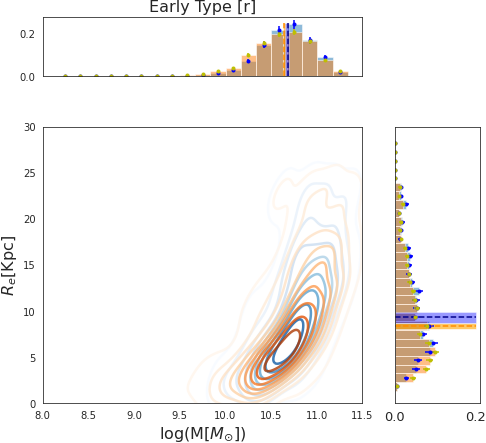}
    \includegraphics[scale=0.34]{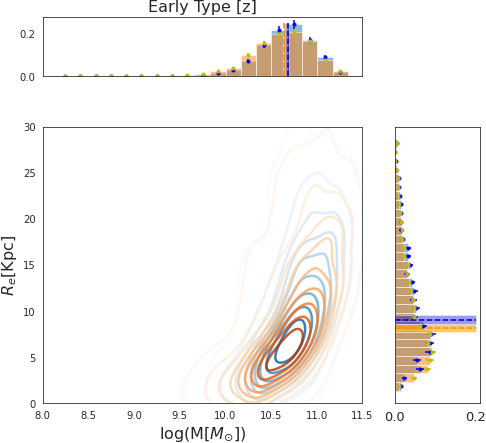}
    \includegraphics[scale=0.34]{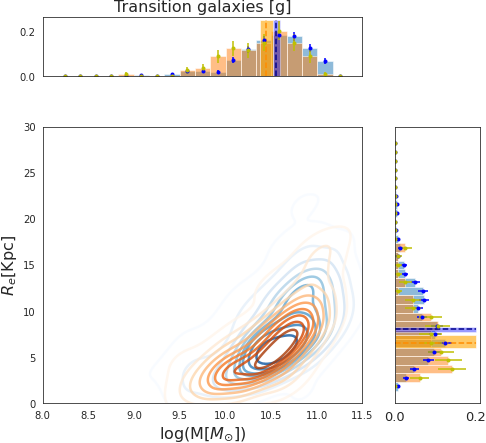}
    \includegraphics[scale=0.34]{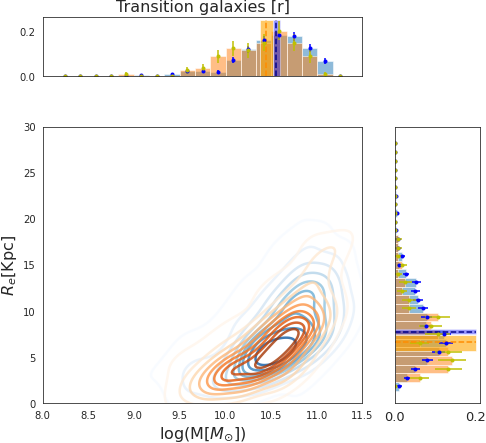}
    \includegraphics[scale=0.34]{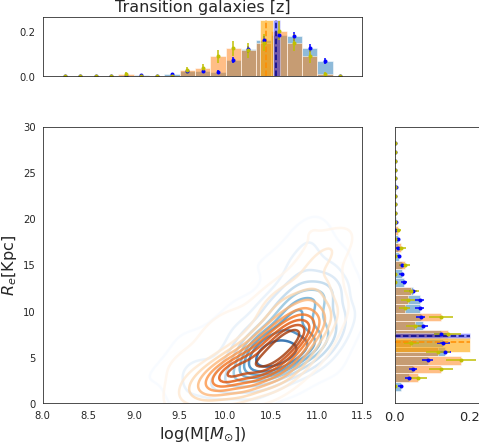}
    \includegraphics[scale=0.34]{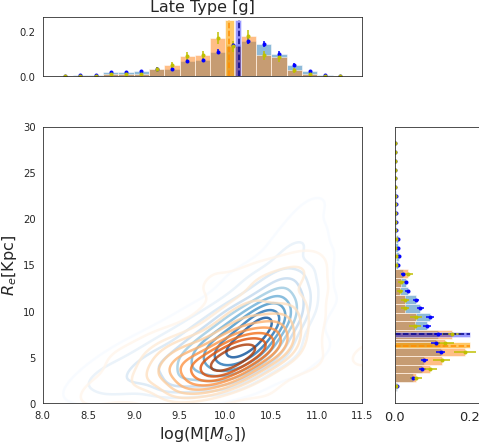}
    \includegraphics[scale=0.34]{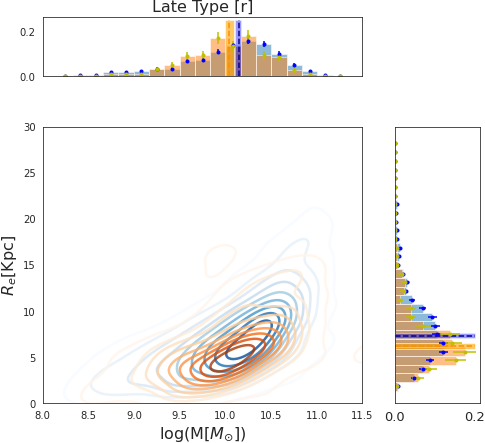}
    \includegraphics[scale=0.34]{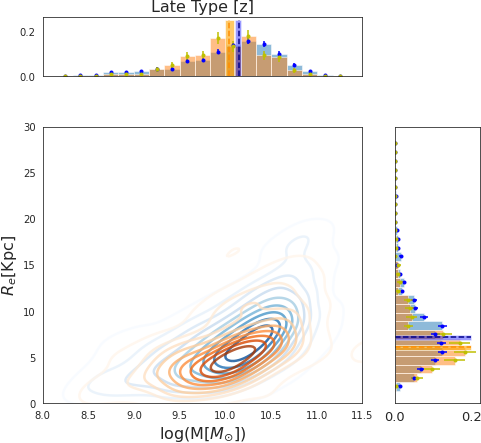}
    \caption{Effective radius as a function of stellar mass for each of the populations; the top three figures for ETGs, the middle three for those in the transition region, and the bottom three for LTGs in $g$, $r$, and $z$ filters from left to right, respectively. Orange contours represent CG galaxies and blue contours are control sample galaxies. At the margin of each plot are histograms of each parameter, in which the dashed lines represent the median and the shaded lines represent the 90\% CI using bootstrapping, and the error bars in the histogram bars are the Poisson counting error.}
    \label{fig:Re_M}
\end{figure*}

In this subsection, we analyze how $R_e$ behaves  as a function of stellar mass, how the structural parameters ($n$ vs. $R_e$) are related for each galaxy population, and how these relations change with the galactic environment. The correlation between galaxy sizes and masses, for nearby galaxies, suggests that the size of a galaxy is an important key to understand its origin (\citealt{2003Conselice}). In Figure \ref{fig:Re_M} we show $R_e$ as a function of stellar mass, for each galaxy population analysed in this work, where blue contours represent galaxies in the control sample and orange contours show galaxies in CGs. The top, middle, and bottom panels show the relation for ETGs, transition galaxies, and LTGs, respectively. The relations are displayed in the $g$, $r$, and $z$-band, from the left to right panels. Marginal to the scatter plot there are histograms of each parameter, where the dashed lines represent the median, the shaded regions represent the 90\% CI using bootstrapping, and the error bars on the histograms are the Poisson counting uncertainty. In general, we find that there are no significant differences in the size-mass relation between the environments in each population where $R_e$ increases as a factor of the $M_*$. However, from the marginal distributions we do observe that the median stellar mass for each population is slightly lower in CGs than in the control sample; the difference is bigger for LTGs. We also find that galaxies appear larger, i.e. bigger $R_e$ in the control sample than in CGs. This is in agreement with what was found by \cite{2012Coenda} i.e., galaxies in CGs are smaller than galaxies in the control sample, but it differs from the findings reported by \cite{Poliakov_2021}, who found that galaxies in CGs are brighter and larger than isolated galaxies.  However, they have deeper images, reaching surface brightnesses of $\sim 28$ mag arcsecond$^{-2}$ in the $r$-band, therefore they can better trace the fainter component of the galaxies, because we achieve surface brightnesses of $\sim 24.5$ mag arcsecond$^{-2}$. In addition, the methodology used by \cite{Poliakov_2021} is different than ours, given that they used GALFIT code, on which GALFITM is based. The main difference is that with GALFIT it is possible to perform a fit just in one filter and not simultaneously as with GALFITM, since they only have data in the $r$-band filter. On the other hand, \cite{2008deng} find no strong dependence on galaxy size and environment, providing a different result on this topic, which was mainly based on a narrow range of luminosity. One of the possible origins for the discrepancy found with the \cite{2008deng} study could be that the method used by them to estimate the size is different than ours: they use the half-light radius direct from the image. 

\begin{figure}
    \centering
    \includegraphics[width=\columnwidth]{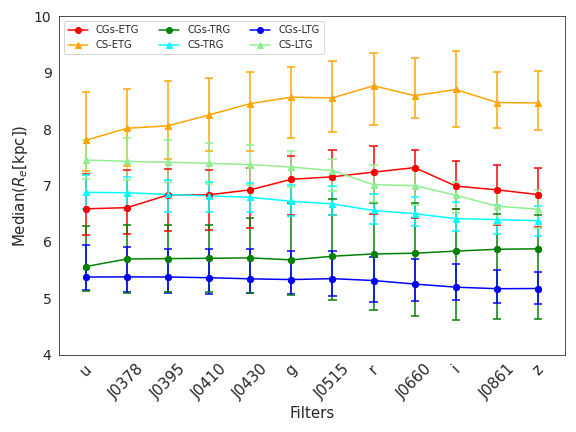}
    \caption{The median effective radius ($R_e$) as a function of S-PLUS filters for ETGs, LTGs, and transition galaxies for in the control sample (triangles) and in the CGs (circles). The error bars are the 90\% confidence interval using bootstrapping.}
    \label{fig:Re_lam}
\end{figure}

In Figure \ref{fig:Re_lam} we show the median effective radius for each galaxy population, as a function of wavelength. The triangles and circles represent the values for the galaxies in the control sample and in CGs, respectively. ETGs, transition galaxies, and LTGs are shown as red/orange, green/light green, and blue/cyan colours, respectively. Uncertainties were estimated from bootstrapping with a CI of 90\%. We find that the median $R_e$ varies with wavelength for all the populations. In the case of ETGs in CGs (red circles), we find that $R_e$ increases with wavelength until  the $J0660$ filter and it begins to decrease from the filters $i$ until $z$. Overall, this parameter has an increase of $4\%$ between the bluest and the reddest filter. For the control sample (orange triangles) the increase in $R_e$ is close to $8\%$ between the bluest and the reddest filter, where $R_e$ increases until the $g$ filter, and after that, it remains almost constant in the median value of $R_e$. For the transition galaxies, $R_e$ increases only $5\%$ in the CGs galaxies (green circles), where the bluest filters remain almost constant in the median value of $R_e$. 
However, the transition galaxies belonging to the control sample (light green triangles) show a decrease in $R_e$ at larger wavelengths, by $13\%$. For the LTGs in the CGs (blue circles), $R_e$ (in median) decreases smoothly, with a difference between the bluest and reddest filter of $4\%$. In the case of the control sample (cyan triangles) the decrease is slightly more pronounced with a difference of $8\%$. 

\begin{figure*}
    \centering
    \includegraphics[scale=0.35]{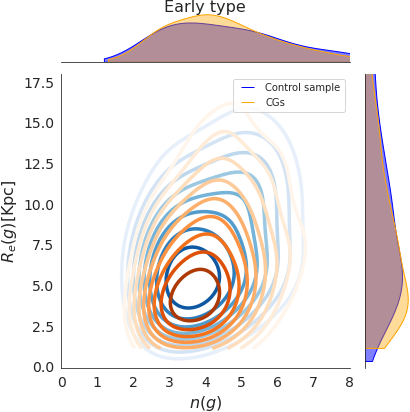}
    \includegraphics[scale=0.35]{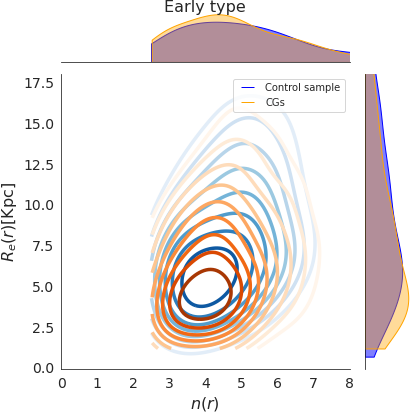}
    \includegraphics[scale=0.35]{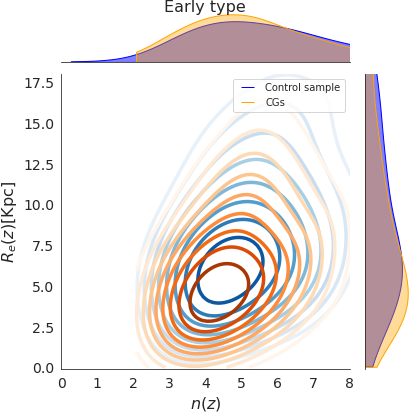}
    \includegraphics[scale=0.35]{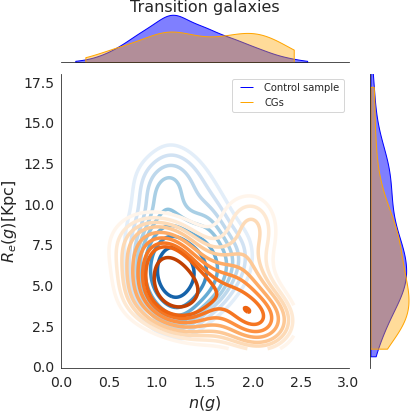}
    \includegraphics[scale=0.35]{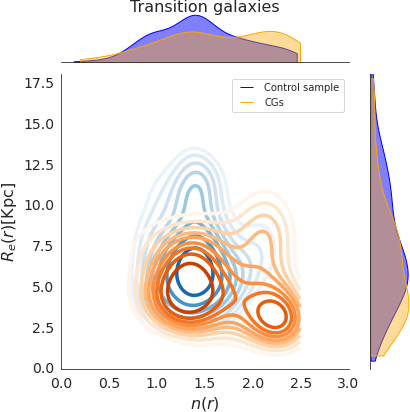}
    \includegraphics[scale=0.35]{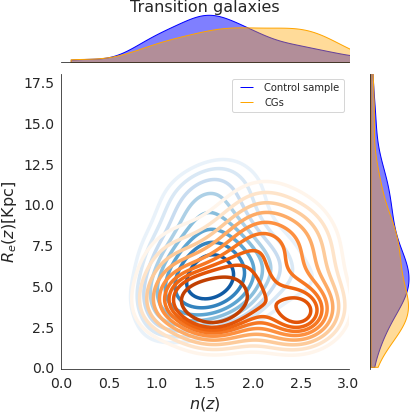}
    \includegraphics[scale=0.35]{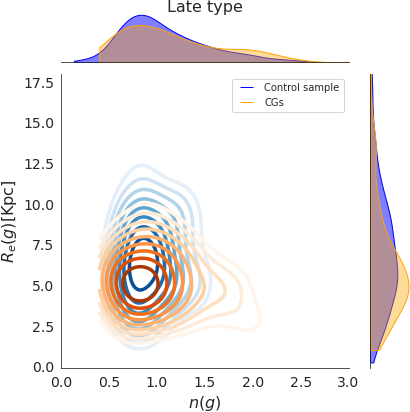}
    \includegraphics[scale=0.35]{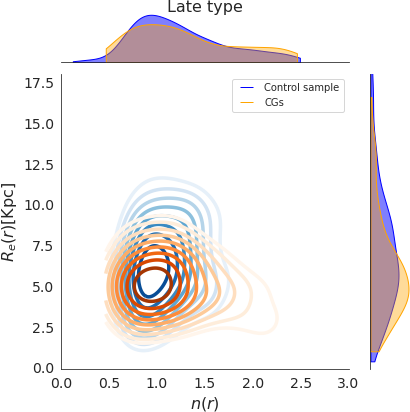}
    \includegraphics[scale=0.35]{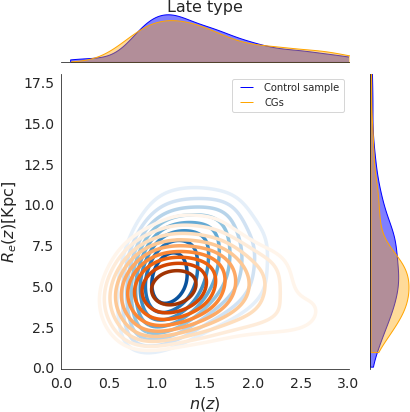}
    \caption{Contours of effective radius as a function of Sérsic index for each of the populations, the top three figures for ETGs, the middle three for those in the transition region, and the bottom three for LTGs, for each environment in three filters $g$,$r$, and $z$. The distribution of each parameter is shown in the margin of each plot.}
    \label{fig:Re_n}
\end{figure*}

In Figure \ref{fig:Re_n} we show the effective radius as a function of the Sérsic index for the filters, $g$, $r$, $z$, from the left to right, where the blue contours represent galaxies in  the control sample and orange contours display the distribution of galaxies in CGs. The top, middle, and bottom panels correspond to ETGs, transition region galaxies, and LTGs, respectively. In general, we find that larger $R_e$ are reached for galaxies in the control sample, as also shown in Figures \ref{fig:Re_M}. In addition, and interestingly, the distribution of transition galaxies in CGs exhibits different behaviour, such that there is a bimodal distribution in this diagram, which is not detected in the control sample. This indicates that transition galaxies in CG have a secondary population of smaller and more compact objects, i.e. lower values of $R_e$ and larger $n$, that we define in this work as a \textit{peculiar galaxy population}. We further discuss this peculiar population in Section \ref{sec:tran_reg}. 

\subsection{Star formation: statistical differences between the CGs and the control sample}
\label{sec:SFR}

\begin{figure*}
    \centering
    \includegraphics[scale=0.40]{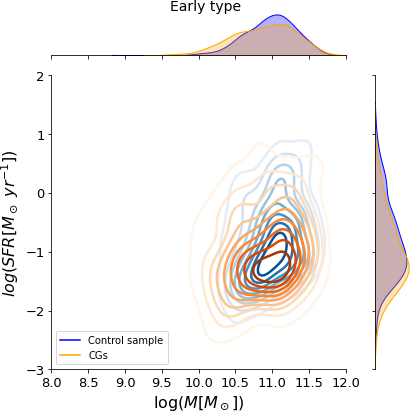}
    \includegraphics[scale=0.40]{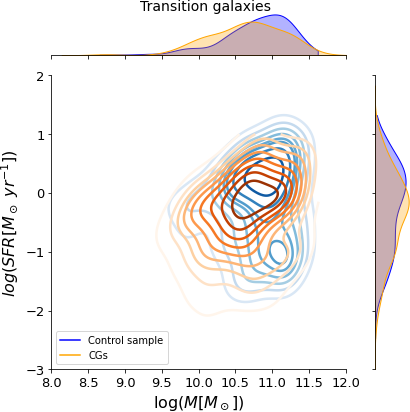}
    \includegraphics[scale=0.40]{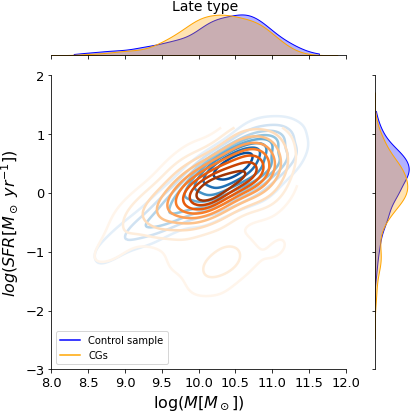}
    \includegraphics[scale=0.40]{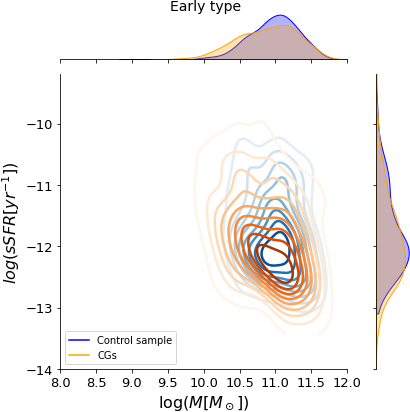}
    \includegraphics[scale=0.40]{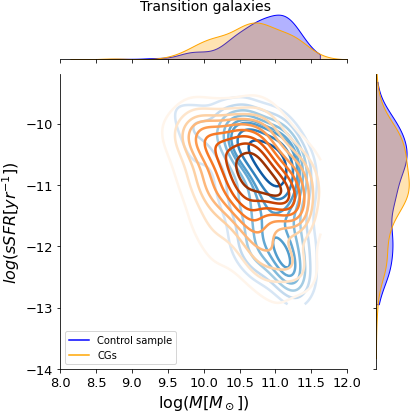}
    \includegraphics[scale=0.40]{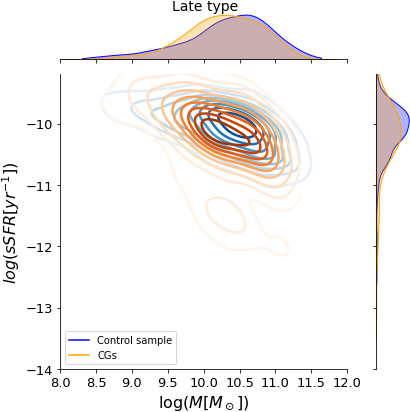}
    \caption{The three top figures display contours of the logarithm of SFR as a function of the logarithm of stellar mass for each of the populations, while the three bottom figures display sSFR instead of SFR: ETGs on the left, transition galaxies on the middle and LTGs on the right. Galaxies in CGs are contours in orange and in the control sample in blue.}
    \label{fig:SFR_SED}
\end{figure*}

In the top panels of Figure \ref{fig:SFR_SED} we show the stellar mass versus SFR for ETGs (left), transition galaxies (middle), and LTGs (right), for galaxies in the control sample (blue contours) and in the CGs (orange contours). We find that, in general, galaxies belonging to CGs have lower SFRs than the galaxies in the control sample. We also observe a bimodality in the distribution of SFR for the transition galaxies in the control sample, whereas the galaxies in the CGs have a distribution that peaks at the middle of this bimodality (at $Log(SFR[M_{\odot} yr^{-1}]) = -11.2$). The bottom panels of Figure \ref{fig:SFR_SED} show the sSFR instated of SFR. We observe that galaxies in CG reach lower values in sSFR than in the control sample. Also, the bimodality previously found in the distribution of $SFR$ as a function of the stellar mass (in the top middle panel of Figure \ref{fig:SFR_SED}) is observed for transition galaxies in the control sample, but we do not find this bimodality in the transition galaxies in CGs. Another important detail to note is that the transition galaxies have intermediate sSFRs compared to the distributions for the LTG and ETG. To quantify these differences, we measured the median sSFR for each population, as shown in Table \ref{tab:resumen}, which is consistent with what we observed in Figures \ref{fig:SFR_SED}. In CGs, the medians of $sSFR$ are lower compared to the control sample. Additionally, based on the criterion proposed by \cite{2013Wetzel}, we observe that CGs contain a larger percentage of quenched galaxies (i.e. $Log(sSFR)\leq-11$), for each galaxy population. This seems to indicate that the CG environment disfavors the star formation activity. To determine whether the differences we find in sSFR between the galaxies in the CGs and the control sample are statistically significant we perform a Kolmogorov-Smirnov (KS) test for the sSFR. We compare the cumulative distribution function (CDF) of this quantity in the CGs and in the control sample and test the null hypothesis that they (i.e. both environments) follow the same distribution. The null hypothesis, in this case, it would mean that, statistically, we cannot be certain of the environmental influence on these populations for star-forming galaxies. We find the null hypothesis to be rejected for ETG and LTG galaxies with $P_{KS}=0.01$ and $0.005$ respectively, while for transition galaxies the null hypothesis cannot be rejected with a $P_{KS}=0.38$. This confirms that the CGs environment may be affecting the star formation process, likely due to tidal interactions, increasing the fraction of quenched galaxies. This scenario will be discussed in more detail in Section \ref{sec:SFR_dis}.

\subsection{Correlation between physical and morphological transformation }
\label{sec:ssfr-mor}

\begin{figure*}
    \centering
    \includegraphics[scale=0.38]{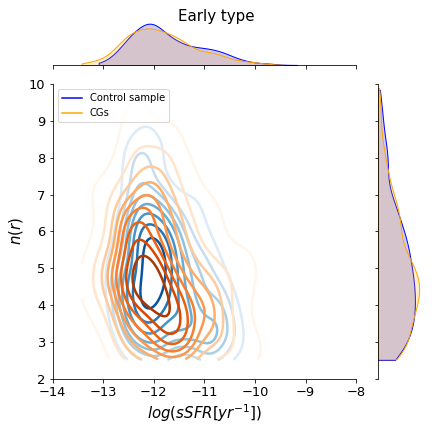}
    \includegraphics[scale=0.38]{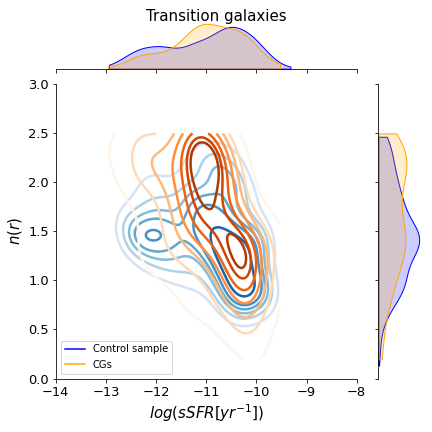}
    \includegraphics[scale=0.38]{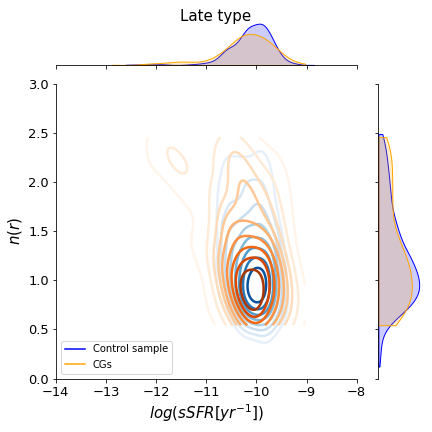}
    \caption{Contours of the Sérsic index in $r$-band  as a function of the logarithm of the sSFR for each of the populations: ETGs on the left, transition galaxies on the middle and LTGs on the right.  Galaxies in CGs are contours in orange and in the control sample in blue.}
    \label{fig:sSFR_n}
\end{figure*}

We find morphological and physical differences when comparing galaxies in CGs with respect to galaxies belonging to the control sample. Thus, it is worth exploring if there is any connection between the morphological and physical parameters of galaxies in both environments, according to their morphological classification. In Figure \ref{fig:sSFR_n}, we show the Sérsic index in the $r$-band as a function of the logarithm of sSFR for the three populations of galaxies: early-type (left plot), transition (middle plot), and late-type galaxies (right plot). The blue contours represent objects in the control sample, while the orange contours represent systems in CGs.

From these plots, we observe that transition galaxies have, on average, higher Sérsic indices and lower sSFR values than LTGs, but higher sSFR values compared to ETGs, indicating their intermediate properties as discussed in the previous section. These results, based on sSFR, are in agreement with our criteria to define early, transition, and late-type galaxies, which was based purely on Sérsic index and colors.

Furthermore, we observe that ETGs and LTGs follow a similar distribution in the $n$-$log(sSFR)$ plane in both environments. However, we find differences when comparing transition galaxies in the control sample with those in CGs. 
While these galaxies in both environments have lower $sSFR$ for higher $n_r$ values; displaying a bimodality in this plane (see Figure~\ref{fig:sSFR_n}), the $n_r$ values are much higher in the CGs for the peak in the bimodality for quenched galaxies (i.e. $log(sSFR)\sim -11$).
We find that a significant fraction of the peculiar galaxies defined in Section \ref{sec:re} contributes to the formation of the larger $n_r$, lower $sSFR$ peak in the bimodality, centered at $n_r\sim2.1$ and $log(sSFR)\sim -11$. This suggests that the peculiar galaxy population is undergoing both physical and morphological transformations.

\section{Discussion}
\label{sec:discussion}

In this section, we discuss the physical meaning of the results found, as well as compare them with other works.

\subsection{Behaviour of the median of Sérsic index and effective radius as a function of wavelength}

Figures \ref{fig:n_fil} and \ref{fig:Re_lam} summarize the dependence of the median value of $n$ and $R_e$ on wavelength. First, we find that galaxies in CGs are smaller, i.e. have lower  $R_e$ values, than those in the control sample when comparing each galaxy population. This could be due to the truncation of galaxies due to tidal interactions that strip material from the outer part of galaxies and enrich the intragroup medium.

For ETGs we observe an increase in both parameters with wavelength in both environments. However, the increase is smoother for galaxies in CGs compared to the control sample. The behavior of these parameters can offer insights into the formation history of ETGs. According to \cite{2009Hopkins}, elliptical galaxies are composed of two components that were generated after a violent relaxation from a major merger of two disk galaxies: a larger, older, and red component, and an inner, younger, and bluer stellar component formed during a central starburst produced during the merger. This scenario could explain why we observe an increase in $n$ and $R_e$ with wavelength.

For LTGs, we find that $n$ increases and $R_e$ decrease as a function of wavelength, regardless of 
 the environment. Therefore, at longer wavelengths, we observe a redder and more concentrated population that is likely associated with the presence of bulges in these galaxies. The variation of these quantities is smooth in both environments, but in CGs the change in $R_e$ is even smoother from the bluer to the redder filters. This, together with the lower values of $R_e$ in CGs, suggests that interactions in this environment are smoothing the drop in the $R_e$ values with wavelength. In other words, the outer more bluer parts of LTGs seen in the control sample is not detected in CGs, due to the truncation of galaxies from the interactions in the latter environment.

For transition galaxies, we see the same behavior as in LTGs, with $n$ increasing for redder wavelengths, with a slightly larger increase in CGs than in the control sample. On the other hand, while $R_e$ decreases for the control sample in this type of galaxies, we find an increase of it for CGs. This indicates that $R_e$ in transition galaxies behaves similar to that of ETGs in CGs, while transition galaxies in the control sample behave in a similar way that LTGs. These results suggest a morphological transformation that occurs in the environment of CGs that does not occur in the control sample of field galaxies. This difference in behavior is related to the population of peculiar galaxies that exist in CGs (see Section \ref{sec:re}), which contain more compact and smaller galaxies. Future studies should pay attention to transition galaxies located in CGs given their importance in galaxy evolution (as we discuss below).

\citet{2010LaBarbera}, \citet{Kelvin2012}, and \citet{2014vulcani} studied the median behavior of $R_e$ and $n$, for a sample of galaxies in various environments; i.e. they did not focus on a particular environment and, instead, were interested in having a large sample of galaxies to analyse their general behavior. \cite{2010LaBarbera} focused on ETGs, while the other two studies classified the galaxies according to their $n$ and their colours. These three investigations find that $R_e$ decreases for ETGs and LTGs, from blue to red bands. Furthermore, they find that the average $n$ increases strongly with wavelength for LTGs, while for ETGs they find that the average $n$ increases smoothly until the $z/Y$ filters and is more stable for longer wavelength, for the $J$, $H$, and $K$ filters $n$ remains constant. \cite{2014vulcani} associate these results for LTGs to the fact that the sample is dominated by two-components bulge-disk galaxies, although they also mention that there is likely to be a contribution from stellar population gradients within each component and from dust attenuation. The behavior for ETGs is interpreted as a superposition of different stellar populations associated with multiple minor merger events, arguing that early-type galaxies are expected to comprise a compact population of red stars formed in situ and a more diffuse, bluer population of stars formed in accreted systems generating their effective radius to decrease with wavelength. It is worth noting that the observations analyzed in \cite{2014vulcani} are shallower than $\sim 25$ mag arcsec$^{-2}$ \citep{Kelvin2012}, thus it is unlikely to observe the diffuse outer regions in these galaxies. In addition, \cite{2021ciria} studied the Hydra cluster and found an increase of $n$ with respect to wavelength for the ETGs ($13\%$ from $u$ to $z$) and a decrease of $n$, by $7\%$ from $u$ to $z$, for the LTGs, which can indicate that star formation is concentrated in the inner regions.

Differences in the way the effective radius ($R_e$) behaves have been observed between ETG galaxies and earlier studies, it is essential to read these results with careful consideration. In this study, our goal is to analyse $R_e$ as a function of wavelength in order to investigate whether the environment of CGs affects this morphological feature compared to a less dense environment, such as field galaxies. Previous studies examined a larger sample of galaxies to explore the relationship between colour and structure within galaxies using optical-near-infrared imaging in bright, low-redshift galaxies, without giving relevance to the environment in which these galaxies are found. The differences with the behavior of the $R_e$ as a function of the wavelength offer a promising opportunity to investigate how the environment impacts $R_e$ changes as a function of wavelength and how this information can be used to further improve our understanding on galaxy formation and evolution, particularly for ETGs. Although this goes beyond the scope of this paper, it highlights the necessity of addressing this issue through hydrodynamical simulations that would enable us to trace the influence of various interactions on galaxies throughout their history, particularly through a multi-wavelength analysis.

\subsection{A peculiar galaxy population in CGs: Are we witnessing morphological transformation in CGs?}
\label{sec:tran_reg}

One of the most interesting results reported in this work is that transition galaxies in CGs display a bimodal distribution in the $R_e-n$ diagram, not seen in the control sample. There is a peculiar population of galaxies in CGs identified as having larger Sérsic index and smaller effective radii than galaxies belonging to the control sample. Indeed, in the marginal plots of $R_e-n$, Figure \ref{fig:Re_n}, the distribution of $n$ presents a clear bimodality in CGs that is not observed in the control sample. A KS test based on the cumulative distribution function (CDF) on this parameter allows us to identify statistically if both samples follow the same distribution. We find that the p-value is $P_{KS}=2.5\times 10^{-4}$, which confirms what is visually observed in Figure \ref{fig:Re_n}, i.e. that transition galaxies from the CG and the control sample do not arise from the same distribution.

This difference found in the population in transition galaxies between CGs and the control sample indicates the existence of a galaxy population that is suffering morphological transformation: Galaxies become more compact and smaller, thus promoting the transformation of transition galaxies into ETG in dense environments. In addition, we have visually inspected all the transition galaxies in CGs. We find that $65\%$ of them are disk-like galaxies, $25\%$ have a spheroidal shape, and $10\%$ are undergoing a merger with a close companion. Since these transition galaxies have colours redder than the LTGs and moderate sSFRs, they are good candidates to fall into the gap found in the medium infrared (MIR) (\citealt{2007AJohnson}, \citealt{2008Gallagher}, \citealt{2010Tzanavaris}, \citealt{2010Walker}, \citealt{2012walker}, \citealt{2013walker}, \citealt{2016Lenkic}). The transition galaxies in CGs appear as a unique sample to study, in detail, how the environment affects galaxy evolution.

\subsection{Physical transformation according to each population type}
\label{sec:SFR_dis}

We find that there are statistical differences in LTG and ETG when comparing the sSFR distribution of galaxies in CGs with the control sample galaxies, while this is not the case for transition galaxies. Additionally, the percentage of quenched galaxies is always higher in the CGs compared to the control sample for all galaxy population types. For LTGs a possible scenario that favors the differences found in sSFR between the two environments are tidal interactions, shocks, and turbulence (\citealt{2015bAlatalo}, \citealt{2016Bitsakis}), causing these gas-rich galaxies to lose material due to interactions in CGs. This is consistent with the scenario proposed by \cite{2001Verdes-Montenegro}, where CG galaxies are H\textsc{i} deficient. For the ETGs there is a higher fraction of quenched galaxies with a lower median sSFR in CGs compared to the control sample.  Furthermore, ETGs have been found preferentially in more dynamically evolved CGs (Montaguth et al. in prep), therefore they have undergone more interactions which explains the differences in the sSFR.

For transition galaxies, which are candidate galaxies to be in the green infrared valley (\citealt{2007AJohnson}, \citealt{2012walker},  \citealt{2016Bitsakis}) we do not find statistical differences in the distribution of sSFR for those galaxies in CGs and the control sample. This result does not imply that they come from the same distribution, it simply means that we do not have enough statistical evidence to support that they are from different distributions. It is worth noting, however, that we do find the median of the sSFR to be lower in CGs compared to the control sample, and the fraction of quenched galaxies to be higher, indicating that there is an environmental effect in CGs that favors galaxies being quenched. Additionally, the transition galaxies in the control sample present a bimodality in the sSFR, not seen in the CGs counterpart: they are either star-forming or quenched, whereas in the CGs we find the galaxies to have intermediate values of sSFR within this bimodality. Note the higher mass of the quenched population in the transition galaxies in the control sample.

\section{Summary and Conclusions}
\label{sec:conclusions}
 
In this work, we used data from 340 compact groups (CGs) in the Stripe 82 region in order to study the evolution of galaxies in dense, low-velocity dispersion environments. In particular, we focus on how this environment affects the morphological and physical properties of galaxies, by comparing to a control sample of isolated galaxies that is analyzed in exactly the same way as our CGs sample. Thus, we have an \textit{homogeneous data set of CG and control sample galaxies}. We used multi-wavelength data from the S-PLUS project, which has 12 filters in the optical. By using this data set and the MegaMorph code we estimated the structural parameters for each galaxy. This information was complemented with GSWL catalogue to obtain the SFR. We divided the galaxy population in each sample as early-type (ETGs), late-type (LTGs), and transition galaxies, according to their Sérsic index ($n$) and colour. The most important findings and results from our analysis are:

\begin{enumerate}
    \item We found that galaxies of all types in CGs have a smaller effective radius than the same type of galaxies located in the control sample. Tidal interactions may favor galaxies losing material in CGs, causing them to have a smaller average effective radius than galaxies in the control sample.
    
    \item We have observed different trends with wavelength in the median values of $n$ and $R_e$ depending on the morphological type and environment. For ETGs, both $n$ and $R_e$ increase with wavelength, with a smoother increase in CGs than in the control sample. This behavior can be related to the formation history of these galaxies, where major mergers contribute to the formation of a red, older, and larger component, along with a younger, bluer, and inner stellar component formed from gas (\citealt{2009Hopkins}). For LTGs, $n$ and $R_e$ increase and decrease, respectively, with wavelength, in both environments, indicative of a redder and more concentrated population, likely associated with the presence of bulges in these galaxies. For transition galaxies, we observe a different behavior in CGs compared to the control sample: the $R_e$ behaves like the ETGs in CGs, while it resembles that of the LTGs in the control sample, suggesting a morphological transformation that occurs in the environment of CGs but not in the control sample.
    
    \item In the $R_e-n$ distribution of the transition galaxies, we observe a bimodality in CGs, that does not appear in the control sample. Therefore, we find a population of CG galaxies that does not follow the same properties that we observe in galaxies located in less dense environments (control sample); we name them "peculiar galaxy population". The galaxies that characterise this sub-sample of transition galaxies in CGs are smaller and more concentrated, which indicates that these galaxies are undergoing a morphological transformation in CGs. 

    \item In CGs, there is a higher fraction of quenched galaxies, regardless of galaxy type, compared to the control sample. Furthermore, there are differences in the distribution of sSFR between LTGs and ETGs when comparing the two environments. On the other hand, in transition galaxies, there is no statistical differences in the sSFR distribution between the two environments analyzed, but there is a higher fraction of quenched galaxies with a lower median sSFR in CGs compared to the control sample. These results suggest that physical processes such as tidal interactions, shocks, and turbulence may play a role in explaining the observed differences in sSFR between CGs and the control sample.

    \item We observe an anti-correlation in the $n_r$-$\log(sSFR)$ plane for transition galaxies, which leads to a bimodality in this plane. We find that a significant fraction of the peculiar galaxy population contributes to the formation of the quenched-galaxies peak in this bimodality, suggesting that the peculiar galaxies undergo not only a morphological but also a physical transformation.
    
\end{enumerate}    

The results presented in this work highlight the importance of studying the morphological and physical properties of galaxies in CGs in order to understand how dense environments affect their evolution. As a follow-up work, and based on these results, we will perform a dynamical analysis of this sample of galaxies to determine whether there is any relation between dynamical properties and the morphological transformation that we find in the transition galaxies in CGs and with the physical properties of each population (ETG, transition galaxies, and LTG) in the CG compared to the control sample (Montaguth et al. in preparation). Additionally, future work with cosmological simulations analyzing compact groups from a multiwavelength approach may be helpful to interpret the different trends that we find for the structural parameters as a function of wavelength. 

\section*{Acknowledgements}

We thank the reviewer, Jon Loveday, for their helpful comments, which improved the quality of this paper. G.P.M acknowledges financial support from ANID/"Beca de Doctorado Nacional"/21202024. 
G.P.M, A.M. and F.A.G. gratefully acknowledge support by the ANID BASAL project FB210003, and funding from the Max Planck Society through a “PartnerGroup” grant. 
G.P.M and A.M acknowledge support by the FONDECYT Regular grant 1212046. F.A.G acknowledge support by the FONDECYT Regular grant 1211370.
ST-F acknowledges the financial support of ULS/DIDULS through a regular project number PR222133. 

The S-PLUS project, including the T80-South robotic telescope and the S-PLUS scientific survey, was founded as a partnership between the Fundação de Amparo à Pesquisa do Estado de São Paulo (FAPESP), the Observatório Nacional (ON), the Federal University of Sergipe (UFS), and the Federal University of Santa Catarina (UFSC), with important financial and practical contributions from other collaborating institutes in Brazil, Chile (Universidad de La Serena), and Spain (Centro de Estudios de Física del Cosmos de Aragón, CEFCA). We further acknowledge financial support from the São Paulo Research Foundation (FAPESP), the Brazilian National Research Council (CNPq), the Coordination for the Improvement of Higher Education Personnel (CAPES), the Carlos Chagas Filho Rio de Janeiro State Research Foundation (FAPERJ), and the Brazilian Innovation Agency (FINEP).

The authors are grateful for the contributions of CTIO staff in helping in the construction, commissioning and maintenance of the T80-South telescope and camera. We are also indebted to Rene Laporte and INPE, as well as Keith Taylor, for their important contributions to the project. We also thank CEFCA staff for their help with T80-South. Specifically, we thank Antonio Marín-Franch for his invaluable contributions in the early phases of the project, David Cristóbal-Hornillos and his team for their help with the installation of the data reduction package jype version 0.9.9, César Íñiguez for providing 2D measurements of the filter transmissions, and all other staff members for their support.
\section*{Data Availability}
 
The data used in this article are published availble. In the case of the photometric data from the S-PLUS survey they are published on the website ($splus.cloud/catalogtools$). The GSWL catalog by \citealt{2018GSWLC} are published on the website ($https://salims.pages.iu.edu/gswlc/$).



\bibliographystyle{mnras}
\bibliography{mnras_template.bib} 


\appendix

\section{Other region}
\label{sec:apen_Other}
\begin{figure*}
    \centering
    \includegraphics[scale=0.34]{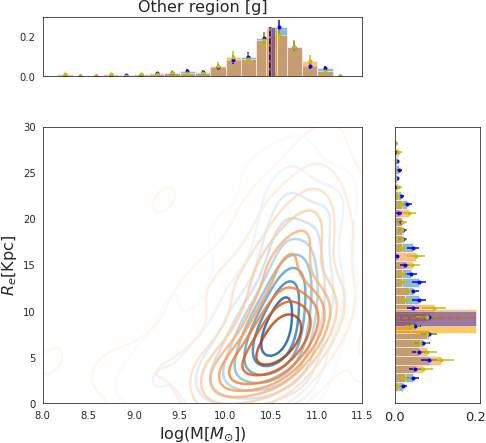}
    \includegraphics[scale=0.34]{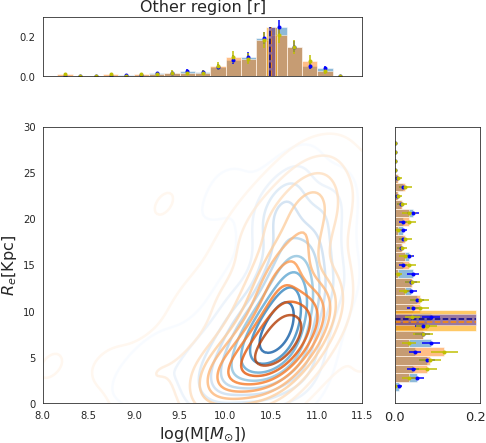}
    \includegraphics[scale=0.34]{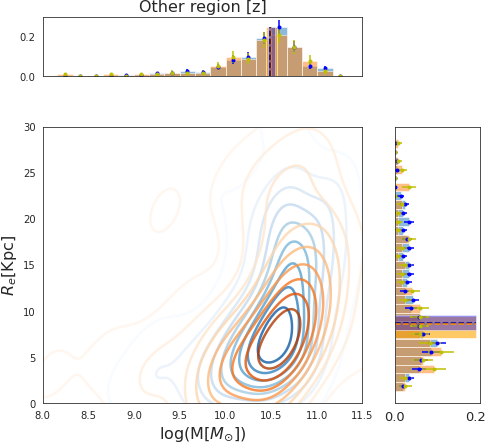}
    \caption{Effective radius as a function of stellar mass for galaxies in the other region, in $g$,$r$ and $z$ filters from left to right of the plot, respectively. Blue contours are control sample galaxies and orange contours represent CG galaxies. At the margin of each plot are histograms of each parameter, in which the dashed lines represent the median and the shaded lines represent the 90\% CI using boostrapping, and the error bars in the histogram bars are the Poisson counting error.}
    \label{fig:R50_o}
\end{figure*}

\begin{figure*}
    \centering
    \includegraphics[scale=0.35]{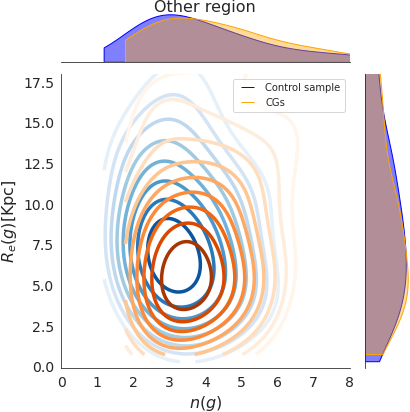}
    \includegraphics[scale=0.35]{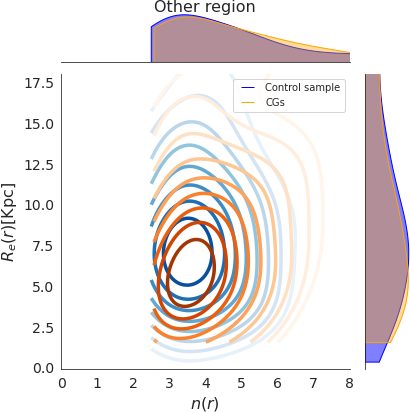}
    \includegraphics[scale=0.35]{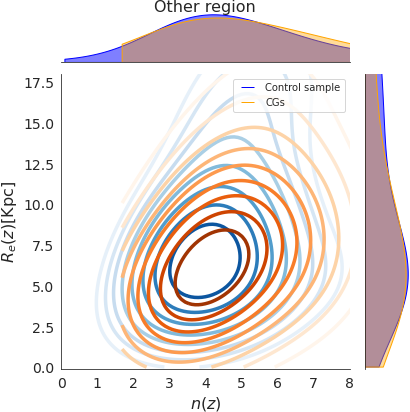}
    \caption{Same as figure \ref{fig:Re_n} but for galaxies in the region that we call other.}
    \label{fig:Re_n_o}
\end{figure*}

\begin{figure*}
    \centering
    \includegraphics[scale=0.35]{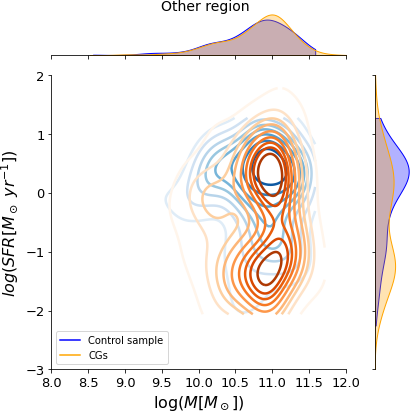}
    \includegraphics[scale=0.35]{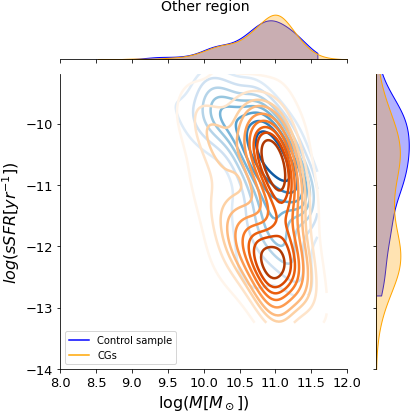}
    \caption{The contours for the SFR (left plot) and $sSFR$ (right plot) as a function of stellar mass for galaxies in the other region. In blue contours for galaxies in the control sample, and in orange contours for CGs. ETGs.}
    \label{fig:SFR_SDSS_other}
\end{figure*}

In the fourth quadrant of the classification proposed by \cite{vika2015megamorph}, and shown in Figure \ref{fig:E_T_population}, we find a population of blue galaxies having high Sérsic indices. We find that $14.9\%$ and $10.4\%$ of CGs and control sample galaxies lie in this region, respectively. In Figure \ref{fig:R50_o} we observe no significant difference between the median effective radius for the environment of each population. Furthermore, the $R_e-n$ distribution for each population (shown in Figure \ref{fig:Re_n_o}) does not show significant variations and reaches similar values to the ETGs in the $R_e$. In this region the mean $Log(sSFR)$ is $-11.2_{-10.9}^{-11.7}$ for the CGs and $-10.6_{-10.5}^{-10.8}$ for the control sample (in a $90\%$ confidence interval, in both cases using SDSS data). In Figure \ref{fig:SFR_SDSS_other} we show the contours for the SFR and $sSFR$ as a function of stellar mass for galaxies in the other region. We observe that the galaxies in CGs span a wider range in the plots compared to the control sample. In particular, we observe a bimodality in galaxies in CGs (lower panels), which could indicate that in this region there is a mixture of different types of galaxies.

For this reason we made a visual inspection of the galaxies in this region. We found that $49\%$ of CGs galaxies have an expected shape; $40\%$ have a disk-like shape and $11\%$ have clear signs of merger. In the case of the control sample, we found a $38\%$ of spheroids and $62\%$ of disk-like objects. This explains why we observe similar $R_e-n$ counterpoints to the ETGs, since spheroidal galaxies would dominate in this plot. In the case of the SFR-mass plots disk-like galaxies dominate, explaining the high values found in the median. This is because $36\%$ of the star-forming galaxies located in CGs, in the fourth quadrant of Figure \ref{fig:E_T_population}, have spheroidal shape or show signs of merging, while in the control sample $22\%$ have spheroidal shape.

\bsp	
\label{lastpage}
\end{document}